\documentclass{article}%
\usepackage{amsfonts}
\usepackage{sw20bams}%
\usepackage{amsmath}%
\usepackage{amssymb}%
\usepackage{graphicx}
\newtheorem{theorem}{Theorem}

\begin{document}

\title{The heat operator in infinite dimensions}
\author{Brian C. Hall\thanks{Supported in part by NSF grant DMS-0555862}\\University of Notre Dame, Dept. of Mathematics, bhall@nd.edu}
\date{October 2007}
\maketitle

\begin{abstract}
Let $(H,B)$ be an abstract Wiener space and let $\mu_{s}$ be the Gaussian
measure on $B$ with variance $s.$ Let $\Delta$ be the Laplacian (\textit{not}
the number operator), that is, a sum of squares of derivatives associated to
an orthonormal basis of $H.$ I will show that the heat operator $\exp
(t\Delta/2)$ is a contraction operator from $L^{2}(B,\mu_{s})$ to $L^{2}%
(B,\mu_{s-t}),$ for all $t<s.$ More generally, the heat operator is a
contraction from $L^{p}(B,\mu_{s})$ to $L^{q}(B,\mu_{s-t})$ for $t<s,$
provided that $p$ and $q$ satisfy%
\[
\frac{p-1}{q-1}\leq\frac{s}{s-t}.
\]
I give two proofs of this result, both very elementary.

\end{abstract}

\section{Introduction}

The heat operator, both on Euclidean space and on Riemannian manifolds, is a
basic tool in finite-dimensional analysis. In infinite-dimensional analysis,
the Laplacian (i.e., the most naive infinite-dimensional generalization of the
finite-dimensional Laplacian) cannot be defined as a self-adjoint operator,
because there is no such thing as Riemannian volume measure in infinite
dimensions. If one replaces the nonexistent volume measure with a Gaussian
measure or something similar on a nonlinear manifold, the Laplacian becomes
not only fails to be self-adjoint but fails to be closable in $L^{2}.$ This
makes it difficult to define the heat operator as a reasonable operator in
$L^{2}.$

In this paper, I will consider only the case of an infinite-dimensional
Euclidean space. In that case, I present one possible way of making the heat
operator into a well-defined, bounded operator.

I thank Professor Leonard Gross for valuable discussions, especially in
pushing me to understand the relationship between the heat semigroup and the
Hermite semigroup in terms of commutation relations.

\section{Gaussian measures}

Let $H$ be an infinite-dimensional, real, separable Hilbert space. Since there
does not exist anything like Lebesgue measure in infinite dimensions, we may
consider instead a Gaussian measure. Let us attempt to construct, then, a
\textquotedblleft standard\textquotedblright\ (i.e., mean zero, variance one)
Gaussian measure on $H.$ This should be given by the nonrigorous expression%
\begin{equation}
d\mu(x)=\frac{1}{Z}e^{-\left\Vert x\right\Vert ^{2}/2}\mathcal{D}%
x,\label{mu.def}%
\end{equation}
where $\mathcal{D}x$ is the nonexistent Lebesgue measure on $H$ and $Z$ is a
normalization constant. Unfortunately, this formal expression does not
correspond to any well-defined measure on $H.$ Specifically, (\ref{mu.def})
\textit{can} be used to assign a \textquotedblleft measure\textquotedblright%
\ to cylinder sets, but this set function does not have a countably additive
extension to the generated $\sigma$-algebra. (A cylinder set is a set that can
be described in terms of finitely many continuous linear functionals on $H.$)

To rectify this situation, we follow the approach of L. Gross in \cite{Gr}. We
introduce a Banach space $B$ together with a continuous embedding of $H$ into
$B$ with dense image. If $B$ is \textquotedblleft enough
bigger\textquotedblright\ than $H,$ in a sense spelled out precisely in
\cite{Gr}, then there is a measure $\mu$ on $B$ that captures the essence of
the formal expression in (\ref{mu.def}). One rigorous way to characterize the
measure $\mu$ is to use (\ref{mu.def}) to define a set function on cylinder
sets \textit{in the larger space} $B.$ This set function, unlike the one on
cylinder sets in $H,$ has a countably additive extension to the generated
$\sigma$-algebra. (This result, of course, assumes Gross's condition on the
embedding of $H$ into $B$.) The original space $H$ turns out to be a set of
$\mu$-measure 0 inside $B.$ The book of Kuo \cite{Ku} is an excellent standard
reference on this material.

A pair $(H,B)$ satisfying Gross's condition is called an \textit{abstract
Wiener space}. The prototypical example is the one in which $H$ is the space
of $H^{1}$ functions on $[0,1],$ equaling $0$ at 0, with inner product given
by%
\[
\left\langle f,g\right\rangle _{H}=\int_{0}^{1}f^{\prime}(x)g^{\prime}(x)~dx.
\]
In this case, one may take $B$ to be the space of continuous functions on
$[0,1]$ equaling 0 at 0. In this case, $\mu$ is the (concrete) Wiener measure,
describing the behavior of Brownian motion.

As an alternative to using (\ref{mu.def}), one can characterize $\mu$ in terms
of its Fourier transform. The measure $\mu$ is the unique one such that for
all continuous linear functionals $\phi$ on $B,$ we have%
\begin{equation}
\int_{B}e^{i\phi(x)}d\mu(x)=e^{-\left\Vert \phi\right\Vert _{H}^{2}%
/2},\label{mu.def2}%
\end{equation}
where $\left\Vert \phi\right\Vert _{H}$ is the norm of $\phi$ as a linear
functional on $H$ (\textit{not} $B$). (That is, $\left\Vert \phi\right\Vert
_{H}$ is the norm of the restriction of $\phi$ to $H.$) In light of the
standard formula for the Fourier transform of a Gaussian, (\ref{mu.def2}) is
formally equivalent to (\ref{mu.def}).

Note from either (\ref{mu.def}) or (\ref{mu.def2}) that it is the geometry of
$H$, rather than the geometry of $B,$ that is controlling the Gaussian measure
$\mu.$ One should think of $\mu$ as the standard Gaussian measure
\textquotedblleft on\textquotedblright\ $H,$ where the larger space $B$ is a
technical necessity, needed to capture the measure.

We can also introduce Gaussian measures with variance $s.$ For any $s>0,$
there is a unique measure $\mu_{s}$ on $B$ such that%
\[
\int_{B}e^{i\phi(x)}d\mu(x)=e^{-\left\Vert \phi\right\Vert _{H}^{2}/2s}
\]
for all continuous linear functionals $\phi$ on $B.$ The measure $\mu_{s}$ is
simply a dilation of $\mu.$

\section{The Laplacian and the heat operator}

If we keep our focus on the geometry of $H$ (rather than $B$), then we can
introduce a Laplacian on $B$ as follows. Let $\left\{  e_{n}\right\}
_{n=1}^{\infty}$ be an orthonormal basis for $H,$ with the property that the
linear functionals $x\rightarrow\left\langle e_{n},x\right\rangle $ extend
continuously from $H$ to $B.$ Then let $\left\{  x_{n}\right\}  _{n=1}%
^{\infty}$ be the coordinate functions (on $B$) associated to this basis; that
is, $x_{n}=\left\langle e_{n},x\right\rangle $ for $x\in B.$ Then we define
the Laplacian to be the operator given by%
\begin{equation}
\Delta=\sum_{n=1}^{\infty}\frac{\partial^{2}}{\partial x_{n}^{2}%
}.\label{delta.def}%
\end{equation}
This operator is defined, for example, on polynomials, that is functions that
can be expressed as polynomials in some finite collection of the $x_{n}$'s.
Note that this operator is really the Laplacian and \textit{not} the
frequently considered number operator (also known as the Ornstein--Uhlenbeck
or Dirichlet form operator). That is, (\ref{delta.def}) is the naive
infinite-dimensional generalization of what is usually called the Laplacian on
$\mathbb{R}^{d}.$ (Those who are emotionally attached to the number operator
should not lose heart; that operator will have its role to play later on.)

We would like to try to define the Laplacian as an unbounded operator in the
Hilbert space $L^{2}(B,\mu)$ or more generally $L^{2}(B,\mu_{s}).$ In fact,
$\Delta$ can be defined densely in $L^{2}(B,\mu_{s}),$ by, for example,
defining it on polynomials. Unfortunately, though, the Laplacian defined in
this way is a nonclosable operator. (This is equivalent to saying that the
adjoint operator is not densely defined. Note that $\Delta$ in not
self-adjoint in $L^{2}$ with respect to a Gaussian measure, even in finite
dimensions.) Nonclosable operators are generally considered to be
pathological; not only does such an operator fail to have a densely defined
adjoint, but it is difficult to make any canonical choice of what its domain
should be.

The nonclosability of the Laplacian can easily be seen by example. Define
functions $f_{n}\in L^{2}(B,\mu_{s})$ by%
\[
f_{n}(x)=\frac{1}{n}\sum_{k=1}^{n}(x_{k}^{2}-s).
\]
Since, as is easily verified, $\left\langle x_{k}^{2}-s,x_{l}^{2}%
-s\right\rangle _{L^{2}(B,\mu_{s})}=0$ for $k\neq l,$ we can see that%
\[
\left\Vert f_{n}\right\Vert _{L^{2}(B,\mu_{s})}^{2}=\frac{c}{n}.
\]
Thus, $f_{n}\rightarrow0$ in $L^{2}(B,\mu_{s})$ as $n$ tends to infinity. On
the other hand, a simple calculation shows that
\[
\Delta f_{n}=2\quad\text{(constant function).}%
\]
Thus, the pair $(0,2)$ is in the closure of the graph of $\Delta,$ which shows
that $\Delta$ is not closable.

Since the Laplacian $\Delta$ is not closable, we cannot expect the heat
operator $e^{t\Delta/2}$ to be any sort of reasonable semigroup in
$L^{2}(B,\mu_{s}).$ One can define $e^{t\Delta/2}$ on polynomials as a
(terminating) power series in $\Delta,$ but the resulting operator is again
not closable. (This shows, in particular, that the heat operator is not
bounded.) The same example functions $f_{n}$ demonstrate the nonclosability of
the heat operator. After all, since $\Delta f_{n}$ is constant, $\Delta
^{2}f_{n}$ is zero and so $e^{t\Delta/2}f_{n}=f_{n}+t\Delta f_{n}/2.$ Thus,
$e^{t\Delta/2}f_{n}$ tends to $t,$ whereas $f_{n}$ tends to zero.

\section{The Segal--Bargmann transform}

The discussion in the previous section shows that we cannot regard the heat
operator $e^{t\Delta/2}$ as a reasonable operator from $L^{2}(B,\mu_{s})$
\textit{to itself}. If, then, we are going to make $e^{t\Delta/2}$ into a
reasonable (preferably bounded) operator, then we must regard it as mapping
from $L^{2}(B,\mu_{s})$ \textit{to some other space}. One way to do this is to
look at the Segal--Bargmann transform. This transform (from one point of view)
consists of applying the heat operator $e^{t\Delta/2}$ to a function and then
analytically continuing the resulting function $e^{t\Delta/2}f$ in the space
variable. Even in the infinite-dimensional case, this makes sense at least on
polynomials. One can then prove an isometry formula that allows one to extend
$e^{t\Delta/2}$ to a bounded operator from $L^{2}(B,\mu_{s})$ into an
appropriate Hilbert space of holomorphic functions on the complexification of
$B.$

\begin{theorem}
Let $\mathcal{P}$ denote the space of polynomials inside $L^{2}(B,\mu_{s}).$
For all $t<2s,$ there is a Gaussian measure $\mu_{s,t}$ on $B_{\mathbb{C}%
}:=B+iB$ such that the map%
\[
f\rightarrow\text{analytic continuation of }e^{t\Delta/2}f,
\]
as defined on polynomials, is isometric from $\mathcal{P}\subset L^{2}%
(B,\mu_{s})$ into $L^{2}(B_{\mathbb{C}},\mu_{s,t}).$
\end{theorem}

This is Theorem 4.3 of \cite{DH}. Here, again, the analytic continuation is in
the space variable (from $B$ to $B_{\mathbb{C}}$) with $t$ fixed. This theorem
shows that $e^{t\Delta/2}$ extends continuously to an isometric map of
$L^{2}(B,\mu_{s})$ into $L^{2}(B_{\mathbb{C}},\mu_{s,t}).$ In \cite{DH}, it is
shown that the image of this extended map is precisely the $L^{2}$ closure of
the holomorphic polynomials in $L^{2}(B_{\mathbb{C}},\mu_{s,t}).$

\section{\textquotedblleft Two wrongs make a right\textquotedblright}

In the preceding section, we saw that the heat operator $e^{t\Delta/2}$,
followed by analytic continuation, can be regarded as a bounded (even
isometric) map of $L^{2}(B,\mu_{s})$ into a Hilbert space of \textquotedblleft
holomorphic\textquotedblright\ functions on $B_{\mathbb{C}},$ provided that
$t<2s.$ (Here \textquotedblleft holomorphic\textquotedblright\ means
\textquotedblleft belonging to the $L^{2}$ closure of holomorphic
polynomials.\textquotedblright) We may ask, however, whether it is possible to
regard the heat operator itself, without the analytic continuation, as a
bounded operator from $L^{2}(B,\mu_{s})$ into some space of functions on $B.$
The answer, as we shall see in this section, is yes. The key is to regard the
heat operator as mapping from $L^{2}(B,\mu_{s})$ into a space defined using a
Gaussian measure with a \textit{different variance}. Specifically, we will see
that for $t<s,$ $e^{t\Delta/2}$ is a bounded operator from $L^{2}(B\,,\mu
_{s})$ into $L^{2}(B,\mu_{s-t}).$

Now, ordinarily, such a \textquotedblleft changing of the
variance\textquotedblright\ is not a good idea. That is, the identity map,
which simply regards a function $f\in L^{2}(B,\mu_{s})$ as an element of
$L^{2}(B,\mu_{s-t}),$ is highly ill defined. After all, it is known that the
measures $\mu_{s}$ and $\mu_{s-t}$ ($0<t<s$) are mutually singular; each
measure is supported on a set that has measure zero with respect to the other
measure. Thus, two functions that are equal $\mu_{s}$-almost every where may
not be equal $\mu_{s-t}$-almost everywhere. Thus, the map $f\rightarrow f$ is
not well defined from $L^{2}(B,\mu_{s})$ to $L^{2}(B,\mu_{s-t}),$ because
elements of $L^{2}$ are not functions but rather equivalence classes of
almost-everywhere equal functions. Alternatively, one can define the identity
map on polynomials (mapping a polynomial in $L^{2}(B,\mu_{s})$ to the same
polynomial in $L^{2}(B,\mu_{s-t})$) and then check that this map is not
closable (consider again the functions $f_{n}$).

We see, then, that the heat operator is not well defined from $L^{2}(B,\mu
_{s})$ to itself and that the identity map is not well defined from
$L^{2}(B,\mu_{s})$ to $L^{2}(B,\mu_{s-t})$. Nevertheless, when we put these
two maps together, two wrongs turn out to make a right: the heat operator is a
well-defined and bounded map from $L^{2}(B,\mu_{s})$ to $L^{2}(B,\mu_{s-t}).$
Somehow, for $f\in L^{2}(B,\mu_{s}),$ if we regard $e^{t\Delta/2}f$ as
belonging to $L^{2}(B,\mu_{s-t})$ rather than $L^{2}(B,\mu_{s}),$ things work
out better. With this point of view, $e^{t\Delta/2}$ actually becomes a
bounded operator.

Actually, more than this can be said. The heat operator is actually bounded
(even contractive) from $L^{p}(B,\mu_{s})$ to $L^{q}(B,\mu_{s-t}),$ for
certain pairs $(p,q)$ with $q>p.$

\begin{theorem}
\label{main.thm}Fix $t<s$ and numbers $p,q>1$ such that%
\[
\frac{q-1}{p-1}\leq\frac{s}{s-t}.
\]

Then the heat operator, initially defined on polynomials, extends to a
contractive operator from $L^{p}(B,\mu_{s})$ to $L^{q}(B,\mu_{s-t}).$
\end{theorem}

Note that since $s$ is greater than $s-t,$ the condition on $p$ and $q$ allows
$q$ to be greater than $p.$ In particular, $p=q$ is always permitted. Theorem
\ref{main.thm} is the main result of this paper. I will present two different
proofs, in the two following sections.

\section{Proof using Hermite polynomials}

Let $\alpha=(\alpha_{1},\alpha_{2},\alpha_{3},\ldots)$ be an infinite
multi-index, in which all but finitely many of the $\alpha_{j}$'s are zero. A
\textbf{polynomial} is then a finite linear combination of functions of the
form $x^{\alpha}:=x_{1}^{\alpha_{1}}x_{2}^{\alpha_{2}}\cdots.$ We let
$\mathcal{P}$ denote the space of all polynomials. Then inside $L^{2}%
(B,\mu_{s})$ we define a \textbf{Hermite polynomial} to be a polynomial of the
form%
\begin{equation}
h_{\alpha,s}(x):=e^{-s\Delta/2}(x^{\alpha}).\label{h.alpha}%
\end{equation}
Here, $e^{-s\Delta/2}$ is the \textit{backward} heat operator, which is
defined on any polynomial by a terminating power series in powers of $\Delta.$
The formula (\ref{h.alpha}) is one of many equivalent ways of defining the
Hermite polynomials. It is known that the Hermite polynomials form an
orthogonal basis for $L^{2}(B,\mu_{s}),$ as $\alpha$ varies over all
multi-indices of the above sort. The normalization is as follows:%
\[
\left\Vert h_{\alpha,s}\right\Vert _{L^{2}(B,\mu_{s})}^{2}=\alpha
!s^{\left\vert \alpha\right\vert },
\]
where $\alpha!=\alpha_{1}!\alpha_{2}!\cdots$Furthermore, for $1<p<\infty,$ the
span of the Hermite polynomials is dense in $L^{p}(B,\mu_{s}).$

We note that for any $t$ and $s$ we have%
\[
e^{t\Delta/2}e^{-s\Delta/2}=e^{-(s-t)\Delta/2},
\]
by the usual power series argument. Thus for $s<t,$ we have%
\[
e^{t\Delta/2}(h_{\alpha,s})=h_{\alpha,s-t}.
\]
This means that the time-$t$ heat operator maps the Hermite polynomials that
go with the Hilbert space $L^{2}(B,\mu_{s})$ to the Hermite polynomials that
go with the Hilbert space $L^{2}(B,\mu_{s-t}).$ This simple observation
provides the first indication that the \textquotedblleft
right\textquotedblright\ way to think of $e^{t\Delta/2}$ is as an operator
from a function space defined using $\mu_{s}$ to a function space defined
using $\mu_{s-t}.$

Actually, $e^{t\Delta/2}$ maps an orthogonal basis for $L^{2}(B,\mu_{s})$ to
an orthogonal basis for $L^{2}(B,\mu_{s-t}).$ Furthermore, since
$(s-t)^{\left\vert \alpha\right\vert }\leq s^{\left\vert \alpha\right\vert },$
it follows easily that $e^{t\Delta/2}$ extends to a contractive mapping of
$L^{2}(B,\mu_{s})$ to $L^{2}(B,\mu_{s-t}).$

To establish the $L^{p}$ to $L^{q}$ properties in Theorem \ref{main.thm}, we
use scaling. We have said that the identity map (i.e., the map $f\rightarrow
f$) is not well defined from $L^{2}(B,\mu_{s})$ to $L^{2}(B,\mu_{s-t}),$ or
vice versa, because the measures $\mu_{s}$ and $\mu_{s-t}$ are mutually
singular. There is, however, a nice map from $L^{2}(B,\mu_{s-t})$ to
$L^{2}(B,\mu_{s}),$ or more generally of $L^{p}(B,\mu_{s-t})$ to $L^{p}%
(B,\mu_{s}),$ consisting of dilation. That is, if we define $D_{s,t}%
:L^{2}(B,\mu_{s-t})\rightarrow L^{2}(B,\mu_{s})$ by%
\[
(D_{s,t}f)(x)=f\left(  \sqrt{\frac{s-t}{s}}x\right)  ,
\]
then this map is well defined and isometric from $L^{p}(B,\mu_{s-t})$ to
$L^{p}(B,\mu_{s}),$ for all $1\leq p\leq\infty$. This amounts to saying that
$\mu_{s}$ can be obtained from $\mu_{s-t}$ by a dilation of $B.$

Meanwhile, how do Hermite polynomials transform under this dilation? Well,
using the formula (\ref{h.alpha}) for the functions $h_{\alpha,s}$, it is not
hard to see that%
\begin{equation}
D_{s,t}(h_{\alpha,s-t})=\left(  \frac{s-t}{s}\right)  ^{\left\vert
\alpha\right\vert /2}h_{\alpha,s}.\label{h.dilate}%
\end{equation}
Now, it makes sense to start with a Hermite polynomial $h_{\alpha,s}\in
L^{p}(B,\mu_{s}),$ apply the heat operator to get $h_{\alpha,s-t}\in
L^{p}(B,\mu_{s-t}),$ and then apply the dilation $D_{s,t}$ to get back to
$L^{p}(B,\mu_{s}).$ We have, by (\ref{h.dilate}),%
\begin{equation}
D_{s,t}e^{t\Delta/2}h_{\alpha,s}=D_{s,t}h_{\alpha,s-t}=\left(  \frac{s-t}%
{s}\right)  ^{\left\vert \alpha\right\vert /2}h_{\alpha,s}.\label{ident0}%
\end{equation}

Let us introduce the \textquotedblleft number operator\textquotedblright%
\ $N_{s}$ defined on polynomials by the condition%
\begin{equation}
N_{s}h_{\alpha,s}=\left\vert \alpha\right\vert h_{\alpha,s}.\label{ns.def}%
\end{equation}
Then (\ref{ident0}) can be rewritten as%
\begin{align}
D_{s,t}\circ e^{t\Delta/2}h_{\alpha,s}  & =\left(  \frac{s-t}{s}\right)
^{N_{s}/2}h_{\alpha,s}\nonumber\\
& =e^{-\tau N_{s}}h_{\alpha,s},\label{ident1}%
\end{align}
where%
\begin{equation}
\tau=\frac{1}{2}\log\left(  \frac{s}{s-t}\right)  .\label{tau.def}%
\end{equation}
Here $e^{-\tau N_{s}}$ is defined on polynomials by setting $e^{-\tau N_{s}%
}h_{\alpha,s}=e^{-\tau\left\vert \alpha\right\vert }h_{\alpha,s},$ in
accordance with (\ref{ns.def}).

Now, every polynomial is a finite linear combination of Hermite polynomials.
What (\ref{ident1}) say, then, is that on the dense subspace $\mathcal{P}$ of
$L^{2}(B,\mu_{s})$ we have%
\begin{equation}
D_{s,t}\circ e^{t\Delta/2}=e^{-\tau N_{s}}.\label{ident2}%
\end{equation}
Note that $e^{-\tau N_{s}}$ is a bounded operator on $\mathcal{P}\subset
L^{2}(B,\mu_{s})$, since its action on the orthogonal basis $\{h_{\alpha}\} $
consists of multiplying by $e^{-\tau\left\vert \alpha\right\vert }.$ Thus
(\ref{ident2}) tells us that $D_{s,t}\circ e^{t\Delta/2}$ has an extension
which is a bounded operator from $L^{2}(B,\mu_{s})$ to itself.

We can, however, say more than this. Nelson \cite{Ne} has shown that $e^{-\tau
N_{s}}$ is contractive from $L^{p}(B,\mu_{s})$ to $L^{q}(B,\mu_{s})$, provided
that%
\begin{equation}
\tau\geq\frac{1}{2}\log\frac{q-1}{p-1}.\label{tau.ineq}%
\end{equation}
More precisely, this can be interpreted as saying that for $\tau$ satisfying
the above condition, $e^{-\tau N}$ has extension from the space $\mathcal{P}$
of polynomials to $L^{p}(B,\mu_{s})$ that maps contractively into $L^{q}%
(B,\mu_{s}).$ (This contractivity property of $e^{-\tau N_{s}}$, where $q>p$
is permitted, is referred to as hypercontractivity.) The condition
(\ref{tau.ineq}) is, in light of (\ref{tau.def}), equivalent to
\begin{equation}
\frac{s}{s-t}\geq\frac{q-1}{p-1}.\label{st.ineq}%
\end{equation}
Thus, whenever (\ref{st.ineq}) holds, $D_{s,t}\circ e^{t\Delta/2}$ has an
extension that is a contractive map of $L^{p}(B,\mu_{s})$ to $L^{q}%
(B,\mu_{s-t}).$ Since $D_{s,t}$ is an isometric isomorphism of $L^{q}%
(B,\mu_{s-t})$ onto $L^{q}(B,\mu_{s}),$ this amounts to saying that
$e^{t\Delta/2}$ has an extension that is a contractive map of $L^{p}(B,\mu
_{s})$ to $L^{q}(B,\mu_{s-t}).$ This establishes Theorem \ref{main.thm}.

\section{The Hermite semigroup and the heat semigroup}

Looking back on the argument in the preceding section, we see that the role of
the Hermite polynomials is not essential. Rather, we used the Hermite
polynomials to obtain the identity (\ref{ident2}), at which point Theorem
\ref{main.thm} is seen to be a consequence of Nelson's hypercontractivity
theorem. That is, our result really hinges on a relationship between the heat
semigroup (the operators $e^{t\Delta/2}$) and the Hermite semigroup (the
operators $e^{-\tau N_{s}}$). What (\ref{ident2}) is saying is that (at least
on polynomials) the Hermite semigroup at time $\tau$ is the same as the heat
semigroup at $t,$ modulo a dilation, where $t$ and $\tau$ are related as in
(\ref{tau.def}). What we want to do in this section is explore two other ways
(besides the Hermite polynomial argument of the previous section) of
understanding this relationship between the two semigroups.

We begin by looking at the integral kernels for the two semigroups. In the
finite-dimensional case, the heat semigroup $e^{t\Delta/2}$ can be computed as
integration against the heat kernel, which is a Gaussian. Meanwhile, in the
finite-dimensional case, the Hermite semigroup $e^{-\tau N_{s}}$ can be
computed using the Mehler kernel, which is also a Gaussian, though of a
slightly more complicated variety. See, for example, the article \cite[Thm.
1]{Sj} of Sj\"{o}gren, which derives the formula for the Mehler kernel in a
way that emphasizes the relationship with the heat kernel. From these formulas
for the kernels, one can easily read off the identity (\ref{ident2}) in any
finite number of variables. This is sufficient to establish (\ref{ident2}) on
polynomials, since each polynomial is a function of only finitely many variables.

As an alternative to using the Hermite polynomial argument of the previous
section or the argument in this section based on formulas for the integral
kernels, we can explore the relationship between the heat semigroup and the
Hermite semigroup using commutation relations. Although we have defined the
number operator by its action on Hermite polynomials ($N_{s}h_{\alpha
,s}=\left\vert \alpha\right\vert h_{\alpha,s}$), $N_{s}$ can also be expressed
as a differential operator, as follows. Let%
\[
D=\sum_{k=1}^{\infty}x_{k}\frac{\partial}{\partial x_{k}}.
\]
Then $Dx^{\alpha}=\left\vert \alpha\right\vert x^{\alpha}.$ From this it
follows, in light of (\ref{h.alpha}), that%
\[
e^{-s\Delta/2}De^{s\Delta/2}h_{\alpha,s}=\left\vert \alpha\right\vert
h_{\alpha,s}.
\]

We now use the standard identity $e^{A}Be^{-A}=e^{\mathrm{ad}_{A}}(B),$ where
$\mathrm{ad}_{A}(B)=[A,B].$ A simple calculation shows that%
\begin{equation}
\lbrack\Delta,D]=2\Delta.\label{relation}%
\end{equation}
Thus if we set%
\begin{align}
N_{s}  & =e^{-s\mathrm{ad}_{\Delta}/2}(D)\nonumber\\
& =D-s\Delta,\label{ns.form}%
\end{align}
where $(\mathrm{ad}_{\Delta})^{n}D=0$ for $n\geq2,$ $N_{s}$ defined in this
way will have the correct behavior on Hermite polynomials.

Our task, then, is to compute
\[
e^{-\tau N_{s}}=\exp\left\{  \tau(s\Delta-D)\right\}  ,
\]
with the aid of the commutation relation (\ref{relation}). We will apply a
special case of the Baker--Campbell--Hausdorff formula. Suppose that $A$ and
$B$ are linear operators on a finite-dimensional vector space and that
$[A,B]=\alpha A.$ Then we have%
\begin{equation}
e^{\tau(A+B)}=e^{\tau B}\exp\left\{  \frac{e^{\tau\alpha}-1}{\alpha}A\right\}
.\label{bch}%
\end{equation}
To prove (\ref{bch}), let $Z(\tau)$ denote the quantity on the right-hand
side. Using the identity
\[
e^{\tau B}Ae^{-\tau B}=e^{\tau\mathrm{ad}_{B}}(A)=e^{-\tau\alpha}A,
\]
it is not hard to show that $Z$ satisfies the differential equation
$dZ/d\tau=(A+B)Z(\tau).$ Since the left-hand side of (\ref{bch}) clearly
satisfies the same differential equation and since the two sides are equal
when $\tau=0,$ we conclude that the two sides are equal for all $\tau.$ (See,
for example, Section 4 of \cite{Di}, where a slight variant of (\ref{bch}) is
analyzed in the context of unbounded operators.)

We wish to apply (\ref{bch}) with $A=s\Delta$ and $B=-D,$ in which case we
would have $\alpha=-2.$ Of course, we should not blindly apply results for
operators on finite-dimensional spaces to operators on infinite-dimensional
spaces. Fortunately, however, there is no problem in this instance, since any
polynomial $p\in\mathcal{P}$ is contained in a finite-dimensional space that
is invariant under both $D$ and $\Delta$ and hence under $N_{s}.$ (A
polynomial $p$, by definition, involves only finitely many monomials
$x^{\alpha}.$ Thus there is some $n$ such that $p$ is contained in the space
of polynomials of degree at most $n$ in some finite collection of variables
$x_{1},\ldots,x_{m}.$)

On each finite-dimensional invariant subspace, then, we have%
\[
e^{-\tau N_{s}}=e^{-\tau D}\exp\left\{  \frac{e^{-2\tau}-1}{(-2)}%
(s\Delta)\right\}  ,
\]
where we may compute using (\ref{tau.def}) that%
\[
-\frac{e^{-2\tau}-1}{2}=-\frac{\frac{s-t}{s}-1}{2}=\frac{t}{2s}.
\]
Thus we get%
\begin{equation}
e^{-\tau N_{s}}=e^{-\tau D}e^{t\Delta/2}.\label{bch2}%
\end{equation}

It remains only to understand the factor of $e^{-\tau D}$ on the right-hand
side of (\ref{bch2}). Recall that $Dx^{\alpha}=\left\vert \alpha\right\vert
x^{\alpha}.$ Thus,
\begin{equation}
e^{-\tau D}x^{\alpha}=e^{-\tau\left\vert \alpha\right\vert }x^{\alpha
}=(e^{-\tau}x)^{\alpha},\label{dilation}%
\end{equation}
where $e^{-\tau}=\sqrt{(s-t)/s}.$ From this we can see that $e^{-\tau
D}=D_{s,t}$ on polynomials and thus (\ref{bch2}) is equivalent (on
polynomials) to the identity (\ref{ident2}).

\section{Concluding remarks}

The boundedness properties of the heat operator given in Theorem
\ref{main.thm} can be deduced from the relationship (\ref{ident2}) between the
heat semigroup and the Hermite semigroup. That relationship, in turn, can be
understood in various ways, using Hermite polynomials, using the integral
kernels, or using commutation relations. In the last approach, the identity
(\ref{ident2}) follows from the commutation relation (\ref{relation}) together
with the special form (\ref{bch}) of the Baker--Campbell--Hausdorff formula.

We have seen in (\ref{dilation}) that $D$ is the generator of dilations. Thus
the commutation relation between $D$ and $\Delta$ in (\ref{relation}) reflects
that the Laplacian transforms in a simple way under dilations. This in turn
reflects that the metric on Euclidean space transforms in a simple way under dilations.

What prospect, then, is there for proving some analog of Theorem
\ref{main.thm} in some other setting, that is, on some (possibly
infinite-dimensional) manifold other than Euclidean space? One possibility is
to consider manifolds where there is some natural sort of dilation operators.
For example, on the Heisenberg group there is a sub-Laplacian that behaves in
a nice way with respect to certain nonisotropic \textquotedblleft
dilations.\textquotedblright\ Thus the Heisenberg group, whether in its
finite- or infinite-dimensional form, is a natural candidate for proving a
theorem similar to Theorem \ref{main.thm}.

On the other hand, even if an infinite-dimensional manifold has no natural
dilations, it is still conceivable that something like Theorem \ref{main.thm}
might hold. Specifically, suppose there exists on an infinite-dimensional
manifold $M$ some natural sort of heat kernel measure $\mu_{s},$ based at a
fixed point in $M.$ (One may think, for example, of path or loop groups and
the heat kernel measures considered by Malliavin \cite{Ma} and Driver--Lohrenz
\cite{DL}, or various infinite-dimensional limits of finite-dimensional groups
and the heat kernel measures of Gordina \cite{Go1,Go2}.) One might hope that
in some cases, the heat operator $e^{t\Delta/2}$ could be a bounded operator
from $L^{p}(M,\mu_{s})$ to $L^{q}(M,\mu_{s-t}),$ for $t<s$ and appropriate
pairs $(p,q).$ One might begin by studying only the Hilbert space case and try
to see whether $e^{t\Delta/2}$ is bounded (or contractive) from $L^{2}%
(M,\mu_{s})$ to $L^{2}(M,\mu_{s-t}).$ Of course, the methods of proof used in
the present paper would not carry over to such a setting. Nevertheless, the
proposed result is of a simple enough form that some other method of proof may
be found.

The conclusion I would like to draw from all of this is that one should not
give up on studying the heat operator associated to the true Laplacian, even
in the infinite-dimensional setting. Here by \textquotedblleft
true\textquotedblright\ Laplacian\ I mean something like the Laplace--Beltrami
operator associated to some Riemannian metric on an infinite-dimensional
manifold, that is the $\nabla^{\ast}\nabla$ operator associated to the
(fictitious) Riemannian volume measure. This is to be contrasted with
something like a number operator or Ornstein--Uhlenbeck operator, which is the
$\nabla^{\ast}\nabla$ operator associated to a Gaussian or heat kernel
measure. Even though the Laplacian is bound to be a pathological operator,
this should not cause us to give up on defining the heat operator. One merely
has to look for the right interpretation of the heat operator, an
interpretation that will allow it to make sense. One candidate for such an
interpretation is to view the time-$t$ heat operator as an operator from
$L^{2}(M,\mu_{s})$ to $L^{2}(M,\mu_{s-t}).$

\end{document}